\documentclass[twocolumn,pra,superscriptaddress,showpacs,amsmath,amssymb]{revtex4}

\usepackage{upgreek, graphicx, subfigure}

\newcommand{\aref}[1]{Appendix~\ref{#1}}

\newcommand{\eref}[1]{Eq.~(\ref{#1})}

\newcommand{\fref}[1]{Fig.~\ref{#1}}

\newcommand{\sref}[1]{section~\ref{#1}}
\newcommand{\Sref}[1]{Section~\ref{#1}}

\newcommand{\ie}{\emph{i.e.}}

\newcommand{\rmd}{~\text{d}}

\newcommand{\tpr}{t^\prime}
\newcommand{\xpr}{x^\prime}
\newcommand{\refl}{\mathfrak{r}}
\newcommand{\trans}{\mathfrak{t}}
\newcommand{\force}{\boldsymbol{F}}
\newcommand{\diffn}{\boldsymbol{D}}
\newcommand{\mst}{\boldsymbol{T}}

\newcommand{\efield}{\boldsymbol{E}}
\newcommand{\bfield}{\boldsymbol{B}}
\newcommand{\efpr}{\boldsymbol{E}^\prime}
\newcommand{\bfpr}{\boldsymbol{B}^\prime}

\newcommand{\im}[1]{\,\text{Im}\!\left\{#1\right\}}
\newcommand{\re}[1]{\,\text{Re}\!\left\{#1\right\}}

\DeclareMathOperator{\sgn}{sgn}

\makeatletter
\newlength \figwidth
\makeatother

\begin{document}

\title{Scattering theory of cooling and heating in opto-mechanical systems}
\author{Andr\'e Xuereb}
\email[To whom all correspondence should be addressed. Electronic address:\ ]{andre.xuereb@soton.ac.uk}
\affiliation{School of Physics and Astronomy, University of Southampton, Southampton SO17~1BJ, United Kingdom}
\author{Peter Domokos}
\affiliation{Research Institute of Solid State Physics and Optics, Hungarian Academy of Sciences, H-1525 Budapest P.O. Box 49, Hungary}
\author{J\'anos Asb\'oth}
\affiliation{Research Institute of Solid State Physics and Optics, Hungarian Academy of Sciences, H-1525 Budapest P.O. Box 49, Hungary}
\affiliation{Leiden Institute of Physics,  NL 2300 RA Leiden, P.O. Box 9504, The Netherlands}
\author{Peter Horak}
\affiliation{Optoelectronics Research Centre, University of Southampton, Southampton SO17~1BJ, United Kingdom}
\author{Tim Freegarde}
\affiliation{School of Physics and Astronomy, University of Southampton, Southampton SO17~1BJ, United Kingdom}

\date{\today}

\pacs{37.10.De, 37.10.Vz, 42.70.Qs}

\begin{abstract}
We present a one-dimensional scattering theory which enables us to describe a wealth of effects arising from the coupling of the motional degree of freedom of scatterers to the electromagnetic field. Multiple scattering to all orders is taken into account. The theory is applied to describe the scheme of a Fabry-Perot resonator with one of its mirrors moving. The friction force, as well as the diffusion, acting on the moving mirror is derived. In the limit of a small reflection coefficient, the same model provides for the description of the mechanical effect of light on an atom moving in front of a mirror.
\end{abstract}

\maketitle

\section{Introduction}
The use of light forces to manipulate mechanical motion has been extended by now from the translational motion of single atoms~\cite{Hansch1975, Wineland1979} to the motional modes of massive systems, such as the oscillations of a micro-mechanical mirror~\cite{Metzger2004, Schliesser2008, Corbitt2007}. The theoretical approach to describe the mechanical effect of light on the center-of-mass motion of atoms is completely distinct from the models dealing with vibrating optical resonators. In the first case, theories are based on the assumption that atoms are very weak scatterers in free space, negligibly perturbing the impinging bright laser beams~\cite{Metcalf2003}. In the other case, the influence of the moving massive component on the radiation field is so strong that it is considered a (moving) boundary condition defining a single or a few modes of the field participating in the opto-mechanical coupling~\cite{Law1995,Braginsky2002}. This is clearly the case for a Fabry-Perot type resonator with one of its mirrors moving~\cite{Gigan2006, Arcizet2006,  Kleckner2006}. We argue that these two cases can be dealt with as two extremes of a general system that can be described in a unified theoretical framework.

In this paper we develop and present a scattering theory for opto-mechanically coupled systems, allowing for the efficient description of the motion of arbitrary combinations of atoms and mirrors interacting through the radiation field. We will restrict the model to one-dimensional motion and small velocities. The main building block is the beamsplitter transfer matrix~\cite{Deutsch1995, Asboth2008}, \ie, the \emph{local relation} between light field amplitudes at the two sides of a scatterer. We will calculate the radiation force acting on a moving scatterer up to linear order in the velocity. The model is completed by including the quantum fluctuations of the radiation force which stem from the quantized nature of the field. We will determine the momentum diffusion coefficient corresponding to the minimum quantum noise level.

The system we will consider in some detail is composed of two mirrors; one of them is fixed in space, whilst the other one is mobile. This is the generic scheme for radiation-pressure cooling of moving mirrors \cite{WilsonRae2007, Marquardt2007, Genes2008}. At the same time, in the limit of low reflection the moving mirror can equally well represent a single atomic dipole interacting with its mirror image in front of a highly reflecting surface~\cite{Eschner2001, Bushev2004, Xuereb2009a}. The scattering model description of this example gives a clear recipe for generalizing the method to more complex systems.

\section{Model}
\begin{figure}[h]
 \centering
 \includegraphics[scale=0.75]{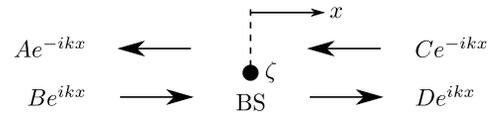}
 \caption{The four different modes that interact through a point-like beamsplitter in 1D.}
 \label{fig:BS}
\end{figure}

Consider a point-like scatterer (or beamsplitter), $\text{BS}$, moving along the `$x$' axis on the trajectory $x_{\text{BS}}(t)$. Outside the scatterer, the electric field  $\efield$ can be expressed in terms of a discrete sum of left- and right-propagating plane wave modes with different wave numbers, $k$, and hence different frequencies, $\omega=kc$:
\begin{equation}
 \label{eq:Efield}
 \efield=\begin{cases}
    \sum_k\big[A(k)e^{-ikx-i\omega t}+B(k)e^{ikx-i\omega t}\big]+\rm{c.c.}\\
    \sum_k\big[C(k)e^{-ikx-i\omega t}+D(k)e^{ikx-i\omega t}\big]+\rm{c.c.}\,,
   \end{cases}
\end{equation}
where $A(k)$ and $B(k)$ are the mode amplitudes on the left side,  $x<x_{\text{BS}}(t)$, while $C(k)$ and $D(k)$ are the amplitudes on the right side, $x>x_{\text{BS}}(t)$, of $\text{BS}$. This is a simplifying assumption and all our results also hold for a continuum of field modes. In accordance, the magnetic field is~\cite{Jackson1998}
\begin{equation}
 \label{eq:Bfield}
 c \bfield=\begin{cases}
    \sum_k\big[-A(k)e^{-ikx-i\omega t}+B(k)e^{ikx-i\omega t}\big]+\rm{c.c.}\\
    \sum_k\big[-C(k)e^{-ikx-i\omega t}+D(k)e^{ikx-i\omega t}\big]+\rm{c.c.}\,.
   \end{cases}
\end{equation}
As depicted schematically in \fref{fig:BS}, the scatterer mixes these waves. Our first goal is the derivation of the transverse matrix $M$ connecting the field amplitudes on the right to those on the left side of a beamsplitter moving at a fixed velocity $v$. This relation is well-known~\cite{Deutsch1995} for an immobile scatterer. Therefore, let us first transform the electromagnetic field into a frame moving with the instantaneous velocity $v$ of the $\text{BS}$.

\subsection{Transfer matrix for an immobile beamsplitter\label{sec:fixedBS}}
In the frame co-moving with $\text{BS}$, the interaction of the field with the scatterer at $\xpr=0$ can be characterized by the single parameter $\zeta$ by means of the one dimensional wave equation~\cite{Jackson1998, Deutsch1995},
\begin{equation*}
\label{eq:WaveEqn}
\left(\partial_{\xpr}^2-\frac{1}{c^2}\partial_{\tpr}^2\right) \efpr(\xpr,\tpr) = \frac{2}{kc^2}\zeta\,\delta(\xpr)\,\partial^2_{\tpr} \efpr(\xpr,\tpr)\,.
\end{equation*}
The electric field can be considered in a modal decomposition similar to \eref{eq:Efield}. Since a fixed beamsplitter couples only the plane waves with identical frequency and wave number, the stationary scattering can be fully described within the closed set of modes
\begin{equation*}
\efpr(\xpr,\tpr) =\begin{cases} A' e^{-ik\xpr -i\omega \tpr} + B' e^{ik\xpr -i \omega \tpr} + c.c. &\xpr < 0\\
C' e^{-ik\xpr -i\omega \tpr} + D' e^{ik\xpr -i \omega \tpr} + c.c. &\xpr > 0\,,
\end{cases}
\end{equation*}
where the index $k$ has been dropped. A  linear relation between the field amplitudes on the right of the scatterer and those on the left can be derived from the wave equation, 
\begin{equation}
\label{eq:TM_fix}
\begin{pmatrix}
 C^\prime\\
 D^\prime
\end{pmatrix}=M_0\begin{pmatrix}
 A^\prime\\
 B^\prime
\end{pmatrix}\,\text{, with}
\end{equation}
\begin{equation}
\label{eq:M0}
M_0 = \begin{bmatrix}
 1-i\zeta & -i\zeta\\
 i\zeta & 1+i\zeta
\end{bmatrix}
= \frac{1}{\trans}
\begin{bmatrix}
 1 & -\refl   \\
\refl   & \trans^2 - \refl^2
\end{bmatrix}\,.
\end{equation}
In the second form of the transfer matrix $M_0$, we expressed it in terms of the reflectivity $\refl$ and transmissivity $\trans$ of the beamsplitter. This latter form is more convenient to describe moving mirrors, while for atoms the scattering strength parameter $\zeta$ can be readily expressed in terms of the polarizability~\cite{Deutsch1995},
\begin{equation*}
 \zeta = \frac{\pi \alpha}{\epsilon_0 \lambda S}\,,
\end{equation*}
where $\alpha$ is the linear polarizability and $S$ is the effective beam cross section. For a two-level, unsaturated atom with transition frequency $\omega_{\text{A}}$ and linewidth $\Gamma$ (HWHM), for example,
\begin{equation}
 \label{eq:ZetaTLA}
 \zeta = \frac{\sigma_{\text{A}}}{2 S} \frac{\Gamma}{\omega_{\text{A}}-\omega - i\Gamma}\,,
\end{equation}
where $\sigma_{\text{A}}= \tfrac{3\lambda^2}{2\pi}$ is the resonant radiative cross section of an atom. In this case the transfer matrix depends on the wave number $k$, which might lead to essential effects, e.g., Doppler cooling, close to resonance with the atom (see \Sref{sec:Force}). 

\subsection{Transfer matrix for a moving beamsplitter\label{sec:MovBS}}
The transformation back into the laboratory-fixed frame involves the change of the coordinates, $\xpr=x-vt$ and $\tpr=t$, and the Lorentz-boost of the electric field up to linear order in $v/c$ \cite[\textsection 11.10]{Jackson1998}:
\begin{equation*}
\efield = \efpr + {v} \bfpr\,,
\end{equation*}
where we assumed that $\efield$ and $\efpr$ are polarized in the `$y$' direction, $\bfield$ and $\bfpr$ are polarized in the `$z$' direction, and the velocity is along the $x$ axis. The electric field in the laboratory frame becomes
\begin{multline*}
\efield(x,t) = \sum_{k'} \Biggl\{A'(k') e^{-ik' (x- vt) -i \omega' t} + B'(k') e^{ik'(x-vt) -i \omega' t} \\
- \frac{v}{c}\Bigl[ A'(k') e^{-ik' (x- vt) -i \omega' t} - B'(k') e^{ik'(x-vt) -i \omega' t}\Bigr]\Biggr\} + \text{c.c.}\\
=\sum_k  \left(1-\tfrac{v}{c}\right) A'\left(k+kv/c\right) e^{-i k (1+v/c) x -i \omega t} \\+ \left(1+\tfrac{v}{c}\right) B'\left(k-kv/c\right) e^{i k (1-v/c) x -i \omega t}\,,
\end{multline*}
which can be expressed as a linear transformation $\hat L(v)$  of the amplitudes,
\begin{equation*}
\begin{pmatrix}
 A(k)\\
 B(k)
\end{pmatrix}=\hat L(-v)\begin{pmatrix}
 A^\prime(k)\\
 B^\prime(k)
\end{pmatrix}\,\text{, with}
\end{equation*}
\begin{equation*}
\label{eq:Lmatrix}
\hat L(v) = \begin{bmatrix}
 \left(1+ \frac{v}{c}\right) \hat{P}_{-v} & 0\\
0 & \left(1- \frac{v}{c}\right) \hat{P}_v
\end{bmatrix}\,.
\end{equation*}
This construction is explored further in~\aref{sec:POper}. Here we defined the operator $\hat{P}_v:f(k)\mapsto f\left(k+k\tfrac{v}{c}\right)$, which represents the Doppler-shift of the plane waves in a moving frame. Obviously, $\hat L^{-1}(v) = \hat L(-v)$ to first order in $\tfrac{v}{c}$. The total action of the moving $\text{BS}$, 
\begin{equation}
\label{eq:ABCD}
\begin{pmatrix}
 C(k)\\
 D(k)
\end{pmatrix}= \hat{M}
\begin{pmatrix}
 A(k)\\
 B(k)
\end{pmatrix}\text{,}
\end{equation}
can then be obtained from
\begin{align}
\label{eq:LML}
 \hat{M} &= \hat L(-v) M_0 \hat L(v)\\
 &=\frac{1}{\trans}\begin{bmatrix}
 1 & -(1-2\tfrac{v}{c})\refl\hat{P}_{2v}\nonumber\\
 (1+2\tfrac{v}{c})\refl\hat{P}_{-2v}& \trans^2-\refl^2
\end{bmatrix}\,,
\end{align}
where we have assumed that $\refl$ and $\trans$ do not depend on the wave number.

Compared to $M_0$ in \eref{eq:M0}, the difference lies in the off-diagonal terms including the Doppler-shift imposed by the reflection on a moving mirror. In other words, the coupled counter-propagating plane wave modes differ in wave numbers, \ie, $k\left(1+\frac{v}{c}\right)$ right-propagating waves couple to $- k\left(1-\frac{v}{c}\right)$ left-propagating waves. Furthermore, if the polarizability itself depends on the wave number $k$, e.g., as in~\eref{eq:ZetaTLA}, the Doppler-shift operator acts also on it. To make this effect explicit, to linear order in~$\tfrac{v}{c}$, $\hat{M}$ can be written as 
\begin{equation*}
\label{eq:Mzeta}
\begin{bmatrix}
 1-i\zeta - i \tfrac{v}{c}{\omega}\tfrac{\partial\zeta}{\partial\omega} & -i\zeta\left[1-\tfrac{v}{c}\big(2 - \tfrac{\omega}{\zeta}\tfrac{\partial\zeta}{\partial\omega}\big)\right]\hat{P}_{2v}\\
 i\zeta\left[1+ \tfrac{v}{c}\big(2 - \tfrac{\omega}{\zeta}\tfrac{\partial\zeta}{\partial\omega}\big)\right]\hat{P}_{-2v}& 1+i\zeta -i\tfrac{v}{c} {\omega}\tfrac{\partial\zeta}{\partial\omega}
\end{bmatrix}\,.
\end{equation*}
The transfer matrix in the laboratory frame can thus be conceived as a 2-by-2 supermatrix acting also in the $k$-space. The amplitude $C$ at a given wave number $k$, \ie, $C(k)$, is combined with the amplitudes $A(k)$ and $B\big(k-2k\tfrac{v}{c}\big)$. A similar statement holds for $D(k)$. 

Starting from the knowledge of the incoming field amplitudes, this transfer matrix allows for calculating the total electromagnetic field around a beamsplitter moving with a fixed velocity. In the next step, we derive the force on the moving scatterer through the Maxwell stress tensor.

\subsection{Force on a medium in an electromagnetic field}

The Maxwell stress tensor (see~\cite[\textsection 6.7]{Jackson1998}) is defined, for a homogeneous medium in one dimension, $x$, as
\begin{equation*}
 \mst=-\frac{\epsilon_0}{2}\Big(\big|\efield\big|^2+{c^2} \big|\bfield\big|^2\Big)\,,
\end{equation*}
where the electric field $\efield$ and the magnetic field $\bfield$, \eref{eq:Efield} and \eref{eq:Bfield}, respectively, have no components along $x$. It is trivial, then, to see that after applying the rotating wave approximation, we obtain
\begin{equation*}
\label{eq:MST}
 \mst_{xx}=-2\epsilon_0\Bigg[\Big|\sum_k A(k)e^{-ikx-i\omega t}\Big|^2+\Big|\sum_k B(k)e^{ikx-i\omega t}\Big|^2\Bigg]\,,
\end{equation*}
since the cross terms in $|\efield|^2$ and $|\bfield|^2$ have opposite signs. Note that $\mst$ varies on time scales of the order of the optical period. Let us now introduce a characteristic time, $\tau$, over which the variations in $\mst$ will be averaged. At $x = 0$, 
\begin{align*}
&\frac{1}{\tau}\int_0^\tau \Big|\sum_k A(k)e^{-i\omega t}\Big|^2\rmd t\\
&\ =\sum_k|A(k)|^2+\sum_{i\neq j}\frac{1}{\tau}\int_0^\tau A(k_i)\big[A(k_j)\big]^\star e^{-i(\omega_i-\omega_j)t}\rmd t\\
&\ \approx \sum_k|A(k)|^2+\sum_{i\neq j}A(k_i)\big[A(k_j)\big]^\star
=\Big|\sum_kA(k)\Big|^2\,.
\end{align*}
In the approximation we assumed that the frequency bandwidth of the excited modes, $\Delta = \max\left\{ \omega_i-\omega_j\right\}$, around the central frequency, $\omega_0$, is so narrow that $\Delta \tau \ll 2 \pi$. Since the broadening is due to the Doppler-shift,  $\Delta \sim 2\omega_0\tfrac{v}{c}$, where $v$ is the speed of the beamsplitter. For example, taking $v$ to be the typical speed of atoms in a magneto-optical trap, we require $\tau\ll \pi/\big(\omega_0\tfrac{v}{c}\big)\sim 10^{-4}~$s. The time needed to reach the stationary regime of scattering is typically much shorter and thus this condition imposed on the averaging time $\tau$ can be safely fulfilled. 

The force on the medium is given by the surface integral of $\mst$ on the surface, $\mathcal{S}$, of a fictitious volume $V=S\,\delta l$ enclosing the medium, where $S$ is the mode area and $\delta l$ the infinitesimal length of the volume along the `$x$' axis. Then, this force is given by
\begin{align}
\label{eq:Forcedef}
 \force&=\oint_\mathcal{S}\mst_{xx} n_x\rmd\mathcal{S}\nonumber\\
&=S\big[\mst_{xx}({x\rightarrow 0^+})-\mst_{xx}({x\rightarrow 0^-})\big]\,,
\end{align}
where $n_x=\sgn(x)$ is the normal to $\mathcal{S}$. Substituting the relevant expressions for $\mst$ into the preceding formula gives
\begin{equation}
\label{eq:MSTForce}
\force=\frac{\hbar\omega}{c}\Big(\big|A\big|^2 +\big|B\big|^2 -\big|C\big|^2-\big|D\big|^2 \Big)\,,
\end{equation}
where $A=[{\hbar\omega/(2S\epsilon_0c)}]^{-1/2}\sum_k A(k)$ is the photo-current amplitude, and similarly for $B$, $C$ and $D$, their modulus square giving the number of photons crossing a unit surface per unit time. Although we considered first the electric field composed of independent modes, in the force expression only the sums of the mode amplitudes occur.

\subsection{Quantum fluctuations of the force}
\label{sec:MSTDiffusion}
In the previous subsection the force was derived based on the assumption that the field amplitudes are c-numbers. In order to describe the inherent quantum fluctuations of the force, we need to resort to the quantum theory of fields and represent the mode amplitudes by operators: $A(k) \rightarrow \hat A(k)$. To leading order the fluctuations of the force acting on a beamsplitter amount to a momentum  diffusion process~\cite{Dalibard1989, Castin1990}. The diffusion coefficient will be evaluated in the following in the minimum, quantum noise limit, which occurs in the case of coherent-state fields~\cite{Glauber1963}.

The diffusion coefficient can be deduced from the  second-order correlation function of the force operator~\cite{CohenTannoudji1992, Gordon1980}
\begin{equation}
\label{eq:StartDiff}
 \big\langle\hat{\force} (t) \hat{\force} (t^\prime) \big\rangle - \big\langle\hat{\force}(t)\big\rangle^2 = \diffn(t) \delta(t -t^\prime)\,.
\end{equation}
The evaluation of this quantum correlation is system specific. Quantum correlations, \ie, the operator algebra of the mode amplitudes $\hat A(k)$, $\hat B(k)$, $\hat C(k)$, and $\hat D(k)$, are influenced by multiple scattering and thus depend on the total transfer matrix of the entire system. The simplest case is a single beam splitter where the ``input'' modes $\hat B(k)$ and $\hat C(k)$ have independent fluctuations. The calculation, delegated to \aref{sec:Diffusion},  includes all the steps needed for the treatment of a general system. The diffusion coefficient for a single beamsplitter is obtained as
\begin{multline}
\label{eq:DiffCoeff}
 \diffn =  (\hbar k)^2 \Bigl(\big|A\big|^2 +\big|B\big|^2+\big|C\big|^2+\big|D\big|^2 \\
 + 2 \re{\refl A^* B -\trans A^* C}\\
+ 2\re{\refl D^*C - \trans D^* B}
  \Bigr)\,,
\end{multline}
where $A, B, C, D$ are the photo-current amplitudes (their modulus square is of the units of 1/sec), obeying \eref{eq:ABCD} for $v=0$.

As an example, let us consider the diffusion coefficient for a two-level atom illuminated by counter-propagating monochromatic light waves.  Using the polarizability $\zeta$, the transmission and reflection coefficients can be expressed as $\trans = 1/(1-i\zeta)$ and $\refl=i\zeta/(1-i\zeta)$, respectively (see~\eref{eq:M0}). \eref{eq:DiffCoeff} can then be rewritten in the form
\begin{equation}
\label{eq:GeneralDiffnBC}
 \diffn =  (\hbar k)^2 \Big[ \tfrac{2 \im{\zeta}}{|1-i\zeta|^2} \big|B-C\big|^2 +  \tfrac{4|\zeta|^2}{|1-i\zeta|^2} \left(\big|B\big|^2 + \big|C\big|^2\right)\Big]\,,
\end{equation}
where the first term, apart from the factor $|1-i\zeta|^2$, corresponds to the result well-known from laser cooling theory, as shown in the next section. Note that the diffusion process due to the recoil accompanying the spontaneous emission of a photon (see~\cite{Gordon1980}) is missing from this result---the detailed modeling of absorption, \ie, scattering photons into the three dimensional space, is not included in our approach.

\subsection{Example: Force on a moving beamsplitter\label{sec:Force}}
We will now use~\eref{eq:MSTForce} to derive a general expression for the force on a moving beamsplitter illuminated by two counterpropagating, monochromatic, plane waves with amplitudes $B_0$ and $C_0$. On using~\eref{eq:ABCD} to express the outgoing field modes in terms of the incoming ones, we note that the outgoing amplitudes comprise two monochromatic terms each:
\begin{equation*}
\label{eq:AD}
A= \frac{i\zeta\left[1-\tfrac{v}{c}\big(2-\tfrac{\omega}{\zeta}\tfrac{\partial\zeta}{\partial\omega}\big)\right] B_0 + C_0}{1-i\zeta\Big(1+\tfrac{v}{c}\tfrac{\omega}{\zeta}\tfrac{\partial\zeta}{\partial\omega}\Big)}\,,
\end{equation*}
and
\begin{equation*}
D= \frac{\Big(1-2i\tfrac{v}{c}\omega\tfrac{\partial\zeta}{\partial\omega}\Big)B_0 + i\zeta\left[1+\tfrac{v}{c}\big(2-\tfrac{\omega}{\zeta}\tfrac{\partial\zeta}{\partial\omega}\big)\right] C_0}{1-i\zeta\Big(1+\tfrac{v}{c}\tfrac{\omega}{\zeta}\tfrac{\partial\zeta}{\partial\omega}\Big)}\,.
\end{equation*}
These relations are substituted into~\eref{eq:MSTForce}, giving
\begin{multline}
\label{eq:ForceTLA}
\force= \bigg\{2\frac{\hbar\omega}{c}\Big/\big|1-i\zeta\big(1+\tfrac{v}{c}\tfrac{\omega}{\zeta}\tfrac{\partial\zeta}{\partial\omega}\big)\big|^2\bigg\}\\
\times\bigg\{\Big(\im{\zeta}+\big|\zeta\big|^2+\tfrac{1}{2}\, \tfrac{v}{c} \omega \tfrac{\partial |\zeta|^2}{\partial\omega} \Big)\Big(\big|B_0\big|^2-\big|C_0\big|^2\Big)\\
- \frac{v}{c} \left(\omega\tfrac{\partial\im{\zeta}}{\partial\omega} - \tfrac{1}{2} \omega \tfrac{\partial |\zeta|^2}{\partial\omega} +2 \big|\zeta\big|^2\right) \Big(\big|B_0\big|^2+\big|C_0\big|^2\Big)\\
+ 2 \left( \tfrac{v}{c} \omega \im{\zeta^\star \tfrac{\partial \zeta}{\partial \omega}} - \re{\zeta}\right) \im{B_0 C_0^\star}\\
+2 \frac{v}{c} \left(2 \im{\zeta} - \omega \tfrac{\partial \im{\zeta}}{\partial\omega} +\tfrac{1}{2} \omega \tfrac{\partial |\zeta|^2}{\partial\omega}\right)\re{B_0C_0^\star} \bigg\}\,.
\end{multline}
For $v=0$ this result reduces to the one in~\cite{Asboth2008}. Most of the $v$-dependent terms arise from the frequency-dependence of the polarisability.  These are the dominant terms in the case of a quasi-resonant excitation of a resonant scatterer, such as a two-level atom, since the prefactor  $\tfrac{\omega}{\zeta}\tfrac{\partial\zeta}{\partial\omega} \sim \tfrac{\omega}{\Gamma}$ expresses resonant enhancement.  The $v$-dependent terms linear in the polarizability $\zeta$ are in perfect agreement with the friction forces known from standard laser cooling theory, both for propagating and for standing waves. For example, assuming identical laser powers from the two sides, giving a standing wave with wavenumber $k_0$, and averaging spatially gives
\begin{equation}
\label{eq:MolassesForce}
\force=-4\hbar k_0^2\big|B_0\big|^2\im{\tfrac{\partial\zeta}{\partial\omega}}v\,,
\end{equation}
for small $\big|\zeta\big|$ and to first order in~$\tfrac{v}{c}$, which can be immediately recognized as the friction force in ordinary Doppler cooling~\cite{Metcalf2003} when one uses the definition of $\zeta$ in~\eref{eq:ZetaTLA}.  Finally, by making similar substitutions into~\eref{eq:GeneralDiffnBC}, we obtain
\begin{equation}
\label{eq:MolassesDiffn}
 \diffn=8(\hbar k_0)^2\im{\zeta}\big|B_0\big|^2\sin^2(k_0 x)\,,
\end{equation}
which, excluding the diffusion effects due to spontaneous emission, matches the standard result in \cite{Gordon1980}. Note, however, that the scattering theory leads to a more general result which is represented by the terms of higher order in $\zeta$. These terms describe the back-action of the scatterer on the field, an effect neglected in free-space laser cooling theory.
\\\ \par
The general result in \eref{eq:ForceTLA} reveals that this velocity-dependent force also acts on a scatterer whose polarisability is independent of the frequency. This is a very general class and we will only focus on such scatterers in the following.

\section{General system of a fixed and a mobile scatterer}

\begin{figure}[htb]
 \centering
 \includegraphics[scale=0.75]{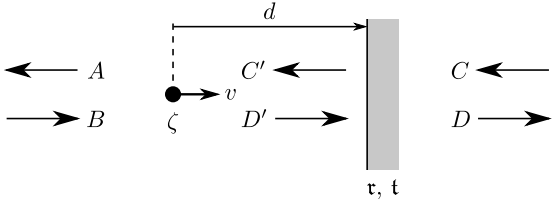}
 \caption{Physical parameters of our model. $A$, $B$, etc. represent the field mode amplitudes.}
 \label{fig:Model}
\end{figure}
Consider the model in~\fref{fig:Model} where the scatterer, or `atom', has a polarizability $\zeta$ uniform over the frequency range of interest. Letting $M_{\text{a}}$, $M_{\text{p}}$ and $M_{\text{m}}$ be the transfer matrices for the atom, propagation and mirror, respectively, we obtain the relation:
\begin{equation*}
\begin{pmatrix}
 A(k)\\
 B(k)
\end{pmatrix} = M_{\text{a}}M_{\text{p}}M_{\text{m}}\begin{pmatrix}
 C(k)\\
 D(k)
\end{pmatrix}\,\text{, where}
\end{equation*}
\begin{align*}
M_{\text{a}} &= \begin{bmatrix}
 1+i\zeta & i\zeta\left(1- 2\tfrac{v}{c}\right)\hat{P}_{2v}\\
 -i\zeta\left(1+2\tfrac{v}{c}\right)\hat{P}_{-2v}& 1-i\zeta
\end{bmatrix} \\
&= \begin{bmatrix}
 M_{11} & M_{12}\hat{P}_{2v}\\
 M_{21}\hat{P}_{-2v} & M_{22}
\end{bmatrix}\,,
\end{align*}
\begin{equation*}
M_{\text{p}} = \begin{bmatrix}
 e^{ikd} & 0\\
 0 & e^{-ikd}
\end{bmatrix}\text{, and }
M_{\text{m}} = \frac{1}{\trans}\begin{bmatrix}
 \trans^2-\refl^2 & \refl\\
 -\refl & 1
\end{bmatrix}\,.
\end{equation*}
The distance between the atom and the mirror is denoted by $d$. Note that the free propagation transfer matrix $M_p$ is non-uniform in the $k$-space, and therefore the Doppler-shift has an influence on the phase shift accumulated between two scattering events.

The boundary condition is set as follows. Since there is no incoming field from the right, $C(k)=0$ for all $k$. The incoming field from the left is assumed to be monochromatic, $B(k)=\mathcal{B}\delta(k-k_0)$, with $k_0$ being the pump wavenumber. The resulting field comprises modes with wavenumbers in a narrow region around $k_0$. In the laboratory frame the field mode $A(k)$ interacts with $B(k-2k\tfrac{v}{c})$ and $C^\prime(k)$ through the Doppler-shift, and similarly for $D^\prime(k)$. From $C(k)=0$ it directly follows that
\begin{align}
\label{eq:InputOutput}
A(k)&=\Big[\refl M_{11}e^{ikd}+M_{12}\hat{P}_{2v}e^{-ikd}\Big]\nonumber\\
&\ \quad\times\Big[\refl M_{21}\hat{P}_{-2v}e^{ikd}+M_{22}e^{-ikd}\Big]^{\!-1}B(k)\nonumber\\
&=\frac{1}{M_{22}}\Big[\refl M_{11}e^{ikd}+M_{12}\hat{P}_{2v}e^{-ikd}\Big]e^{ikd}\nonumber\\
&\ \quad\times\sum_{n=0}^{\infty}\left(-\refl\frac{M_{21}}{M_{22}}\right)^{\!n} e^{2in kd \left[1-(n+1)\tfrac{v}{c}\right]}\nonumber\\
&\ \quad\quad\quad\quad\times B\big(k-2nk\tfrac{v}{c}\big)\,.
\end{align}
We will need the sum of amplitudes, $\mathcal{A}=\int A(k)\rmd k / \mathcal{B}$, defined relative to the incoming amplitude $\mathcal{B}=\int B(k) dk$. Note that $\int \hat{P}_vf(k)\rmd k=\int f(k)\rmd k$. Thus, to first order in $\tfrac{v}{c}$,
\begin{align}
\label{eq:IntA}
\mathcal{A}&=\frac{M_{12}}{M_{22}}+\left(\frac{M_{12}}{M_{22}}-\frac{M_{11}}{M_{21}}\right)\nonumber\\
&\ \quad\times\ \sum_{n=1}^{\infty}\left(-\refl\frac{M_{21}}{M_{22}}\right)^{\!n} \left[1+2in(n-1)k_0 d\tfrac{v}{c}\right]e^{2in k_0d}\,.
\end{align}

It is worth introducing the reference point at a distance $L =2 N \pi/k_0$ from the fixed mirror, where the integer $N$ is such that the moving atom's position $x$ is within a wavelength of this reference point. Then the atom--mirror distance can be replaced by $d=L-x$, and $k_0 L$ drops from all the trigonometric functions. The solution, \eref{eq:IntA}, has a clear physical meaning, in that the reflected field, $\mathcal{A}$, can be decomposed into an interfering sum of fields: the first term is the reflection directly from the atom, whereas the summation is over the electric field undergoing successive atom--mirror round-trips. We can also write the preceding expression in closed form:
\begin{multline}
\label{eq:ReflectedField}
\mathcal{A} =\frac{1}{1-i\zeta}\Biggl\{ i\zeta + \refl \frac{e^{-2 i k_0x}}{1 - i\zeta -\refl i\zeta e^{-2 i k_0x}}\\
- 2 i\frac{v}{c} \zeta \Biggl[1 - \frac{\refl^2 e^{-4 i k_0x}}{\big(1 - i\zeta -\refl i\zeta e^{-2 i k_0x}\big)^2} \\
-2i k_0(L-x) \frac{\refl^2 (1-i\zeta) e^{-4 i k_0x}}{\big(1 - i\zeta -\refl i\zeta e^{-2 i k_0x}\big)^3}\Biggr] \Biggr\}\,.
\end{multline}
This result is valid for arbitrary $\zeta$. The main virtue of our approach is clearly seen, in that we can smoothly move from $\zeta=0$, which indicates the absence of the mobile scatterer, to $|\zeta|\rightarrow\infty$, which corresponds to a perfectly reflecting mirror, \ie, a moving boundary condition.

Let us outline some of the generic features of the above calculation that would be encountered in a general configuration of scatterers. By using the formal Doppler-shift operators, we benefit from the transfer matrix method in keeping the description of the system as a whole within two-by-two matrices. The input-output relation for the total system is always obtained in a form similar to that of \eref{eq:InputOutput}. As long as the Doppler broadening is well below the transient time broadening of the system, the calculation of forces and diffusion requires solely the sum of the mode amplitudes. An important point is that the integrated action of the Doppler-shift operator $\hat P_v$ on monochromatic fields is a shift in $k$-space. Therefore, by interchanging the order of terms and putting the $\hat P_v$ terms just to the left of the input field amplitudes, they can be eliminated, such as in \eref{eq:IntA}. Finally, up to first order in $v/c$, the resulting power series, a trace of multiple reflections, can be evaluated in a closed form, as shown in \eref{eq:ReflectedField}. In conclusion, the illustrated method lends itself for the description of more complex schemes, for example, the cooling of a moving, partially reflective mirror in a high-finesse Fabry-Perot resonator \cite{Bhattacharya2008}.

\subsection{Force acting on the mobile scatterer}
\begin{figure}[tbp]
\subfigure[]{
   \centering
   \includegraphics[width=\figwidth]{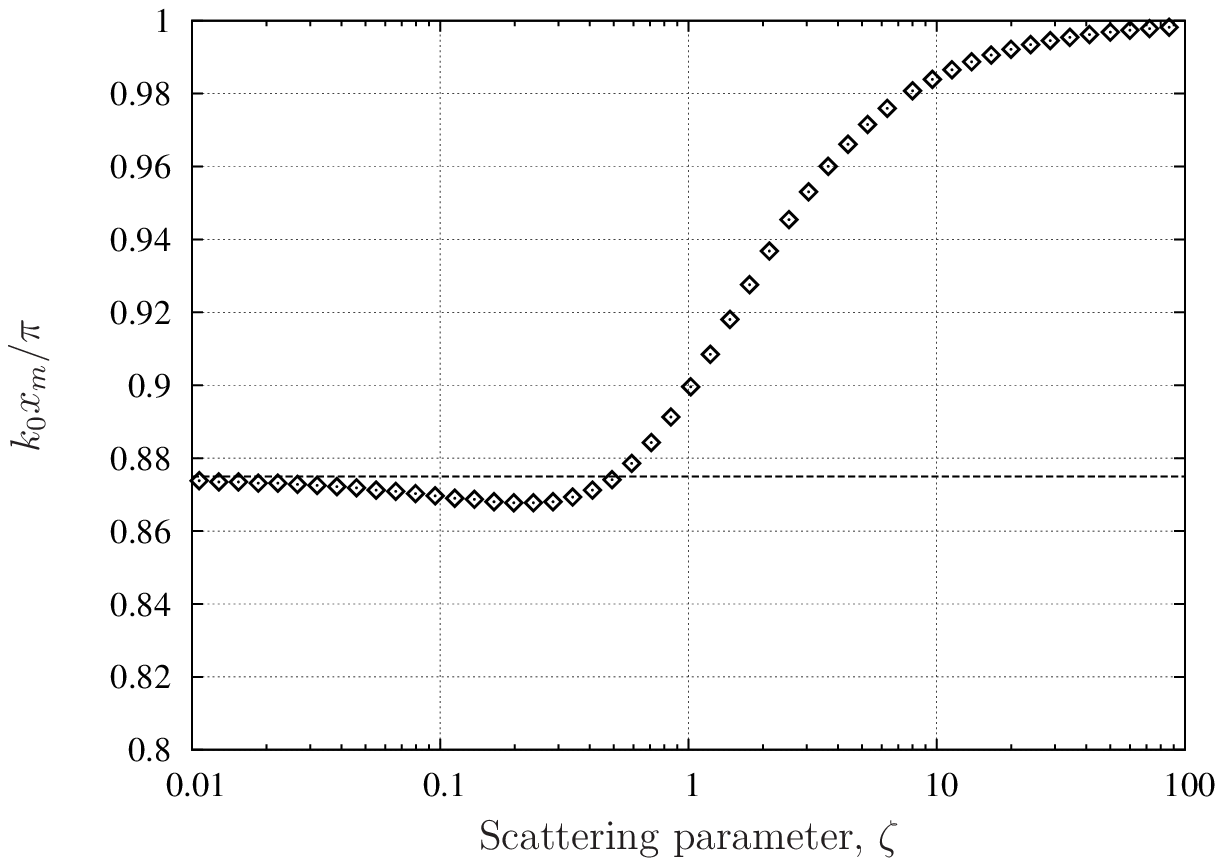}
   \label{fig:Phase}
}
\subfigure[]{
   \centering
   \includegraphics[width=\figwidth]{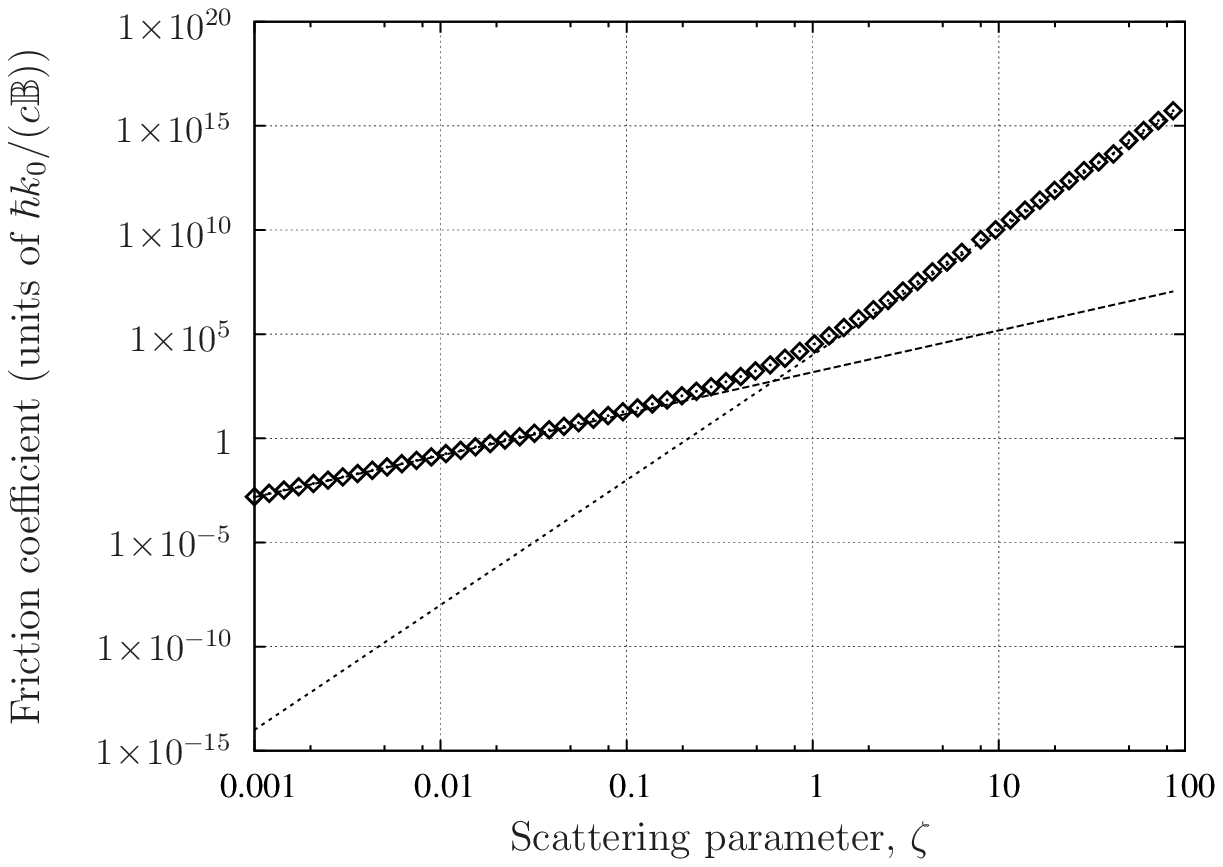}
   \label{fig:Maxfric}
}
\caption{(a) The position of the maximum friction force, $k_0x_m$, as a function of the dimensionless scattering parameter $\zeta$ (on a semilog scale) acting on the scatterer in \fref{fig:Model}, with the fixed mirror being a perfect mirror. This position shifts from $7\pi/8$ to $\pi$ on increasing $\zeta$. (b) A similar plot, showing the maximum friction force as a function of $\zeta$ (on a log-log scale) with $k_0L=100$. In the limit of small $\zeta$, the force scales as $\zeta^2$ (cf.~\eref{eq:MirrorCoolForce}; dashed line) whereas in the limit of large zeta it scales as $\zeta^6$ (cf.~\eref{eq:FPCoolForce}; dotted line).}
\end{figure}

To obtain the force on the moving scatterer, we also need to evaluate $C^{\prime}(k)$ and $D^{\prime}(k)$:
\begin{align}
\label{eq:Cprime}
\begin{pmatrix}
 C^{\prime}(k)\\
 D^{\prime}(k)
\end{pmatrix} &= \begin{bmatrix}
 1-i\zeta &  -i\zeta \left(1 - 2 \tfrac{v}{c}\right)\hat{P}_{2v}\\
 i\zeta \left(1 + 2 \tfrac{v}{c}\right)\hat{P}_{2v}^{-1} & 1+i\zeta
\end{bmatrix} \begin{pmatrix}
 A(k)\\
 B(k)
\end{pmatrix}\,,
\end{align}
where we applied the inverse of the transfer matrix $M_\text{a}$. Next, we make the following definitions:
\begin{align*}
 \mathbb{A}&=|\mathcal{A}|^2\text{, }\quad \mathbb{B}=|\mathcal{B}|^2\text{,}\\
 \mathbb{C}&=\frac{1}{\mathbb{B}}\left|\int C^{\prime}(k)\rmd k\right|^2\text{,}\quad
 \mathbb{D}=\frac{1}{\mathbb{B}}\left|\int D^{\prime}(k)\rmd k\right|^2\,,
\end{align*}
and a simple calculation leads to
\begin{align*}
 \mathbb{C}=&\left|1-i\zeta\right|^2\mathbb{A}+\left|i\zeta\big(1-2\tfrac{v}{c}\big)\right|^2\nonumber\\
&+2\re{i\zeta^{\star}(1-i\zeta)\big(1-2\tfrac{v}{c}\big)\mathcal{A}}\text{,}\\
 \mathbb{D}=&\left|i\zeta\big(1+2\tfrac{v}{c}\big)\right|^2\mathbb{A}+\left|1+i\zeta\right|^2\nonumber\\
&+2\re{i\zeta(1+i\zeta^{\star})\big(1+2\tfrac{v}{c}\big)\mathcal{A}}\,.
\end{align*}
Thereby the force acting on the scatterer is obtained as 
\begin{align}
\label{eq:GeneralForce}
 \force &= (\hbar\omega/c)\mathbb{B}\left(\mathbb{A}+1-\mathbb{C}-\mathbb{D}\right)\nonumber\\
 &= -2 {\hbar k_0 \mathbb{B}} \Big(\left[ | \zeta|^2 \left(1 + 2 \tfrac{v}{c}\right) +\im{\zeta} \right] \mathbb{A}\nonumber\\
 &\phantom{=  -2 {\hbar k_0 \mathbb{B}}\ \Big(}+ |\zeta|^2 \left(1 - 2 \tfrac{v}{c}\right) - \im{\zeta}\nonumber\\
&\phantom{=  -2 {\hbar k_0 \mathbb{B}}\ \Big(}+ 2\re{i\zeta (1-i\zeta) \mathcal{A}}\Big)\,,
\end{align}
where $\mathcal{A}$ has to be substituted from \eref{eq:ReflectedField}.
\begin{figure}[tbp]
   \centering
   \includegraphics[width=\figwidth]{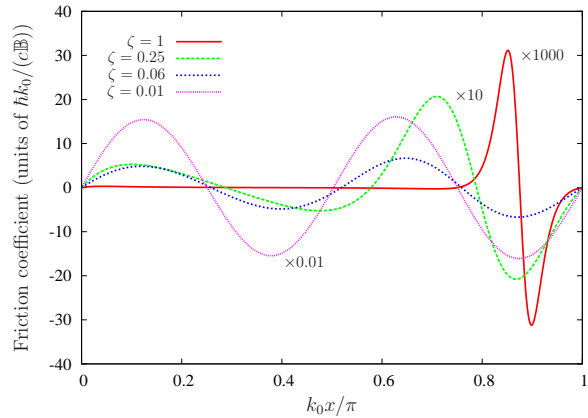}
   \caption{(Color online.) The position dependence of the linear coefficient of the velocity-dependent force acting on the mobile scatterer in \fref{fig:Model}, for various scattering parameters $\zeta$, evaluated by using \eref{eq:ReflectedField} and \eref{eq:GeneralForce} with $k_0 L=100$. The fixed mirror is assumed to be a perfect mirror. In order to fit all the curves into the same range, they are divided by the factors indicated in the figure.}
   \label{fig:Friction}
\end{figure}
The coefficient of the term linear in velocity, the `friction coefficient' $\beta$,  is plotted in \fref{fig:Friction} as a function of the position $x$ in a half-wavelength range for various values of $\zeta$. When varying the coupling strength from $\zeta=0.01$ up to $\zeta=1$, the friction coefficient transforms between two characteristic regimes. For small coupling the linear velocity dependence tends to a simple sinusoidal function while, for large coupling, the friction exhibits a pronounced resonance in a narrow range. This resonance arises from the increased number of reflections between the mobile scatterer and the fixed mirror. It can be observed that the resonance shifts towards $k_0 x =\pi$ on increasing $\zeta$. In the opposite limit of small $\zeta$,  the maximum friction is obtained periodically at $\big(n-\tfrac{1}{4}\big)\pi/2$ according to the sinusoidal function. The position of the maximum friction is plotted in \fref{fig:Phase}, showing the transition from $7\pi/8$ to $\pi$. The maximum friction force is plotted in \fref{fig:Maxfric}, showing the two limiting cases of $\zeta^2$ behavior, in the limit of small $\zeta$, and $\zeta^6$ behavior, in the limit of large $\zeta$. These two cases are described in \sref{sec:MovingAtom} and \sref{sec:MovingMirror}, respectively.

\subsection{Diffusion coefficient}

The calculation of the diffusion coefficient proceeds along the same lines as that corresponding to a single beamsplitter, shown in \aref{sec:Diffusion}. The difference is that the modes $B(k)$ and $C'(k)$ around the mobile scatterer are not independent, for the reflection at the fixed mirror mixes them. Therefore, all the modes $A$, $B$, $C'$, and $D'$ have to be expressed in terms of the leftmost and rightmost incoming modes, $B(k)$ and $C(k)$, respectively. Instead of the derivation of such a general result for the diffusion, here we will restrict ourselves to the special case of $\refl=-1$ ($\Leftrightarrow$ perfect mirror) and real $\zeta$ ($\Leftrightarrow$ no absorption in the moving mirror). In this special case the diffusion calculation simplifies a lot, because (i) the perfect mirror prevents the modes $C$ from penetrating into the interaction region, and (ii) quantum noise accompanying absorption does not intrude in the motion of the scatterer. 

Only the modes $\hat B(k)$ impart independent quantum fluctuations. When all the amplitudes around the scatterer are expressed in terms of $\hat B(k)$, and are inserted into the force correlation function given in \eref{eq:StartDiff}, the commutator $[\hat b(t),\hat b^\dagger(t')]$ appears in all the terms (see \aref{sec:Diffusion}). Straightforward algebra leads to
\begin{equation}
\label{eq:GeneralDiff}
\diffn=\hbar^2 k_0 ^2\mathbb{B}(\mathbb{A} +1-\mathbb{C}-\mathbb{D})^2\,.
\end{equation}
We emphasize that the above result is not general: the diffusion is not necessarily proportional to the square of the force. This simple relation here follows from the assumptions, $\refl=-1$ and $\im{\zeta}=0$, declared above. 

To be consistent with the calculation of the friction force linear in velocity, the diffusion should be evaluated only for $v=0$. From the ratio of these two coefficients,  the steady-state temperature can be deduced. The velocity independent components of the modes obey the following relations: $\mathbb{A} = 1$ and $\mathbb{C'} = \mathbb{D'}$ (all incoming power is reflected). Therefore the diffusion coefficent further simplifies,
\begin{equation}
\diffn = 4 \hbar^2 k_0 ^2\mathbb{B} \Bigg(1-\frac{1}{\big|1 - i\zeta + i\zeta e^{-2 i k_0x}\big|^2}\Bigg)^{\!\!2}\,.
\end{equation}

\begin{figure}[tbp]
   \centering
   \includegraphics[width=\figwidth]{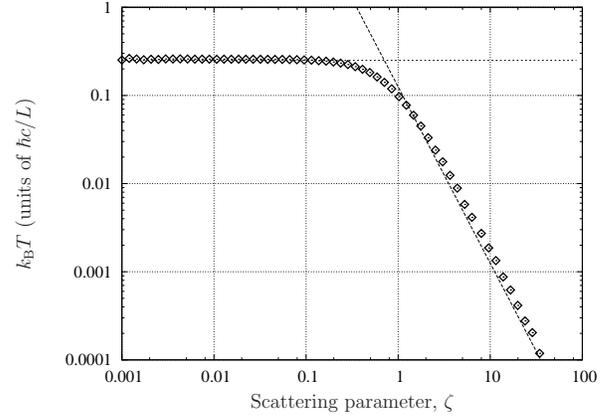}
   \caption{Characteristic temperature for the two-scatterer system of \fref{fig:Model}, given by the ratio of the diffusion and friction coefficients in the points where the friction is maximum, as a function of the dimensionless scattering parameter $\zeta$ on a log-log scale. Constant and $1/\zeta^2$ dependence can be read off in the limits of small and large $\zeta$, respectively. The fixed mirror is a perfect mirror.}
   \label{fig:Temperature}
\end{figure}

In \fref{fig:Temperature}, the temperature $k_B T= \diffn/(2 \beta)$, where $\beta$ is the friction coefficient, is plotted as a function of the scattering parameter $\zeta$. The friction and the diffusion coefficients are taken at the position where the friction is maximum, as shown in \fref{fig:Phase}.
The two limits of small and large scattering parameter $\zeta$ will be analysed in \sref{sec:MovingAtom} and \sref{sec:MovingMirror}, respectively.

\section{Atom in front of a perfect mirror}
\label{sec:MovingAtom}

An atom pumped with a far off-resonance beam can be modelled as a moving mirror with small and real $\zeta$. In this section we accordingly truncate our expressions to second order in $\zeta$. We also assume that the fixed mirror is perfect; \ie, $\refl=-1$ and $\trans=0$. Thus,
\begin{align}
\label{eq:Force}
 \force=2\hbar k_0\mathbb{B}\big\{&2\zeta\im{\mathcal{A}}-2\zeta^2\re{\mathcal{A}}\nonumber\\
&-\zeta^2\big(1+\tfrac{v}{c}\big)\mathbb{A}-\zeta^2\big(1-\tfrac{v}{c}\big)\big\}\,.
\end{align}
To obtain $\force$ to second order in $\zeta$, we need $\mathcal{A}$ to first order. Using~\eref{eq:IntA} and~\eref{eq:Force}, we obtain:
\begin{align}
\label{eq:CurlyA}
 \mathcal{A}=&-e^{-2i k_0x}+\zeta(i-2ie^{-2i k_0x}+ie^{-4i k_0x})\nonumber\\
&+\zeta\tfrac{v}{c}\big[-2i+2ie^{-4i k_0x}-4 k_0  (L-x) e^{-4i k_0 x}\big]\,,
\end{align}
and
\begin{align}
\label{eq:MirrorCoolForce}
 \force=4\hbar k_0\mathbb{B}\big(&\zeta\sin(2 k_0x)\nonumber\\
&-\zeta^2\big\{2\sin^2( k_0x)\big[4\cos^2( k_0x)-1\big]\big\}\nonumber\\
&-\zeta^2\tfrac{v}{c}\big[4\sin^2(2 k_0x)\nonumber\\
&\phantom{-\zeta^2\tfrac{v}{c}\big[}\ -4 k_0 (L-x) \sin(4 k_0x)\big]\big)\,,
\end{align}
in agreement with~\cite{Xuereb2009a}. In the far field ($x\gg\lambda$), the dominant friction term in the preceding expression is the last term, which renders the $\sin(4 k_0x)$ position dependence shown in \fref{fig:Friction} for $\zeta=0.01$.
\par
We are now in a position to derive the diffusion coefficient for this system. By substituting~\eref{eq:CurlyA} into~\eref{eq:GeneralDiff} and setting $v=0$, we obtain
\begin{equation*}
 \diffn=8(\hbar k_0)^2\zeta^2\mathbb{B}\,.
\end{equation*}
This allows us to estimate the equilibrium temperature for such a system at a position of maximum friction:
\begin{equation}
\label{eq:MMCTemp}
 k_{\text{B}}T\approx\frac{\hbar}{2\tau}\text{, where }\tau=2(L-x)/c\,,
\end{equation}
which we note is identical in form to the Doppler temperature for a two-level atom undergoing free-space laser cooling~\cite{Metcalf2003}, but where we have replaced the upper state lifetime, $1/(2\Gamma)$, by the round-trip time delay between the atom and the mirror. Note that this temperature corresponds to the constant value presented in \fref{fig:Temperature} for $\zeta<0.1$.

\section{Optical resonator with mobile mirror\label{sec:Mirrorcooling}}
\label{sec:MovingMirror}
After the small polarizability case of the previous section, we will now consider the $|\zeta| \rightarrow \infty$ limit. We again assume that the fixed mirror of the resonator is perfect, with $\refl=-1$, and that $C=0$. For simplicity, we assume that the moving mirror has a real polarizability. We expand the field mode amplitudes as power series in $v/c$, such that $\mathcal{A}=\mathcal{A}_0+\tfrac{v}{c}\mathcal{A}_1+\dots$, and similarly for $\mathcal{C}^{\prime}$.

Let us first calculate the field in the resonator for $v=0$. It follows from \eref{eq:Cprime} that
\begin{equation*}
\mathcal{C}_0^\prime = (1-i \zeta) \mathcal{A}_0 -i\zeta = -\frac{e^{-2i\varphi}}{1-i\zeta+i\zeta e^{-2 i \varphi}}\,,
\end{equation*}
with $\varphi = k_0 d$, which has a maximum at $\varphi_0$ obeying
\begin{equation*}
\tan(2\varphi_0) = - \frac{1}{\zeta}\,.
\end{equation*}
In the limit of $\zeta\rightarrow \infty$, the resonance is Lorentzian:
\begin{equation*}
\mathcal{C}_0^\prime = -\frac{e^{-2i\varphi}}{2 i (1-i\zeta) \left[(\varphi-\varphi_0) - i \tfrac{1}{4 \zeta^2}  \right]}\,,
\end{equation*}
with a width of $1/(4\zeta^2)$.

The perfect mirror reflects the total power incoming from the left, $\mathbb{B}$.  Moreover, for real $\zeta$, there is no absorption in the moving mirror, so the outgoing intensity has to be equal to the incoming one: $\mathbb{A} = 1$. This is true if $v=0$; for $v\neq 0$, the field can do work on the mirror. The expansion of the back reflected intensity  to linear order  in velocity reads $\mathbb{A} = 1 + 2\tfrac{v}{c} \re{\mathcal{A}_0^*\mathcal{A}_1}$.  Extracting the velocity-dependent terms for the general form of the force in \eref{eq:GeneralForce}, it reduces to 
\begin{equation*}
\force_1= \tfrac{v}{c} 4 \hbar k_0 \mathbb{B} \zeta\im{\mathcal{A}_1\big/\left(1 + i \zeta -i \zeta e^{2i\varphi}\right)}\,,
\end{equation*}
which, after some algebra, leads to
\begin{equation}
\label{eq:FPCoolForce}
\force_1 = - \tfrac{1}{2} \tfrac{v}{c} \hbar k_0^2 L \frac{(\varphi-\varphi_0)}{\zeta^4\, \left[\left( \frac{1}{4\zeta^2}\right)^2 + (\varphi-\varphi_0)^2 \right]^3} \mathbb{B}\,.
\end{equation}
On substituting $\kappa = c/\big(4L\zeta^2\big)$, $\Delta_C = - c(\varphi-\varphi_0)/L$, $\eta^2/(2\kappa)=\mathbb{B}$, and $G=c^2k_0^2/L^2$, the friction force renders that derived from the usual radiation pressure Hamiltonian in \aref{sec:RadPressCool}. 

Expressing the field modes interacting with the mobile mirror in terms of the input field mode and performing a calculation similar to that leading to \eref{eq:DiffCoeff} readily gives
\begin{equation*}
\diffn \approx 4(\hbar k_0)^2 |\mathcal{C}^\prime_0|^4 \mathbb{B} \approx \frac{(\hbar k_0)^2\mathbb{B}}{ 4\zeta^4 \left[ \left( \frac{1}{4\zeta^2}\right)^2 + (\varphi-\varphi_0)^2\right]^2}\,.
\end{equation*}
The resulting temperature thereby attains a minimum at $4 \zeta^2 (\varphi-\varphi_0)=1$, \ie, $\Delta_C=-\kappa$, in analogy with free-space Doppler cooling, at which point we have
\begin{equation}
\label{eq:MirrorCoolingTemp}
k_{\text{B}} T \approx  \frac{\hbar c}{8 \zeta^2 L} = \tfrac{1}{2} \hbar \kappa\,.
\end{equation}
Again, this asymptotic behavior is reflected in \fref{fig:Temperature} for large $\zeta$. We note the similarity of the preceding expression with the temperature of an atom cooled in a cavity, in the good cavity limit~\cite{Horak2001}. We conjecture that this is due to the fact that both systems can be considered to involve the coupling of a laser with a system having a decay rate $\kappa$. This result also holds for the case of an atom undergoing mirror-mediated cooling, as can be seen in \eref{eq:MMCTemp}.
\par
It is also important to note that the above discussion only treats the effects of the light fields on the scatterer. As such, the temperature limit, \eref{eq:MirrorCoolingTemp}, is intrinsic to the light forces, and the mechanical damping and heating processes present in a real, macroscopic mirror-cooling setup are not taken into account. In practice, these heating processes may dominate over the heating induced by the quantum noise in the light field \cite{Saulson1990, Cohadon1999}. In such cases, radiation pressure cooling is a possible means to lower the equilibrium temperature owing to the additional, optical damping process.

\section{Conclusions}
We have presented a powerful extension of an existing theoretical framework to analyse the interaction between light and matter. The theory we presented is based on the transfer matrix method for dealing with the interaction between scatterers and a light field, and is therefore able to handle complex optical systems, made from several elements, with relative ease. Through the use of the Maxwell stress tensor one can calculate the force acting on any of the elements in the system. We have generalized the transfer matrix for slowly moving scatterers, thereby the corrections first-order in $v/c$ can be calculated for the electromagnetic field as well as for the radiation force acting on the scatterer.  Furthermore, one can express this force in terms of the operators representing the quantized field modes interacting with the scatterer and consequently derive the momentum diffusion of the scatterer due to the quantum noise present in the fields. Our scattering theory can also transparently cover the whole range of interaction strengths, from the perturbative interaction between a weak standing wave and a single atom to the very strong (quasi boundary-condition) interaction between a pump light field and a Fabry-Perot cavity with a moving mirror.
\par
We also applied this framework to three different laser cooling configurations: optical molasses, mirror-mediated cooling and cooling of micromirrors. We derived the forces on an atom arising from its interaction with the light field, as well as an estimate for the equilibrium temperature an ensemble of atoms is expected to reach through this interaction. In the case of optical molasses, which corresponds to the well-known Doppler temperature limit, the theory provides for additional force and diffusion terms related to the effect of the back-action of the atom on the radiation field. Although for single atoms in free space this back-action is feeble, it is responsible for the modification of equilibrium properties \cite{Weidemuller1998, Asboth2007b} and for collective effects in large optical lattices \cite{Asboth2007a}. In the latter cases of a moving scatterer in front of a fixed mirror, our results are valid for arbitrary scattering strength, \ie, spanning the parameter range from a single atom to high-reflectivity mirror.

\begin{acknowledgments}
This work was supported by the UK Engineering and Physical Sciences Research Council (EPSRC) grant EP/E058949/1, by the \emph{Cavity-Mediated Molecular Cooling} working group within the EuroQUAM programme of the European Science Foundation (ESF) and by the National Scientific Research Fund of Hungary (NF68736, T049234).
\end{acknowledgments}

\appendix

\section{The Doppler-shift operator\label{sec:POper}}
\begin{figure}[h]
 \centering
 \includegraphics[scale=0.75]{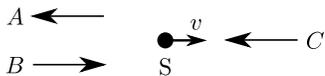}
 \caption{Reflection and transmission of a moving scatterer. $B$ and $C$ are the input field modes, and $A$ is the output field mode. A further output field mode (`$D$') is not drawn because it is not relevant to our discussion in this section.}
 \label{fig:POper}
\end{figure}

Consider the situation in~\fref{fig:POper}, in the laboratory frame, where $\text{S}$ is a scatterer and suppose that $B$ and $C$ are known. $A(k)$ has contributions arising from both $B\big(k+2k\tfrac{v}{c}\big)$ and $C(k)$, where $k$ is any arbitrary wave number, written separately as:
\begin{align*}
 A_B(k)&=a_1 B(k+2k\tfrac{v}{c})\,\text{, and}\\
 A_C(k)&=a_2 C(k)\,.
\end{align*}
We can therefore express $A(k)$ as
\begin{equation*}
 A(k)=a_1 B(k+2k\tfrac{v}{c})+a_2 C(k)\,.
\end{equation*}
Defining $\hat{P}_v$ by $\hat{P}_v:f(k)\mapsto f(k+k\tfrac{v}{c})$, we have
\begin{equation*}
 A(k)=\hat{P}_{2v}a_1 B(k)+a_2 C(k)\,.
\end{equation*}
A similar expression, involving $\hat{P}_v^{-1}=\hat{P}_{-v}$, holds for $D(k)$. These two operators can then be introduced into~\eref{eq:TM_fix} as part of the Lorentz transformation, and thus into the transfer matrix for the moving scatterer, giving rise to the form shown in~\eref{eq:LML}. The resulting transformation, for the transfer matrix $M$, of a scatterer moving with velocity $v$ can be written as:
\begin{equation*}
\begin{bmatrix}
(1-\tfrac{v}{c})\hat{P}_v & 0\\
0 & (1+\tfrac{v}{c})\hat{P}_v^{-1}
\end{bmatrix}M\begin{bmatrix}
(1+\tfrac{v}{c})\hat{P}_v^{-1} & 0\\
0 & (1-\tfrac{v}{c})\hat{P}_v
\end{bmatrix}\,,
\end{equation*}
to first order in $\tfrac{v}{c}$, where the ordering of the elements of $M$ is as described in the text. Note that this relation is general, in the sense that the elements of $M$ can depend on $k$ (see~\sref{sec:Force}).

For any finite $v$, $\hat{P}_v$ is trivially a bounded operator, having unit norm. This property follows from the important relation $\int \hat{P}^m_v f(k)\rmd k = \int f(k)\rmd k\text{,}$ for any function $f(k)$ and any integer $m$.

This operation can be generalized to $n=2,3$ dimensions. We define a new operator by $\hat{S}_i(\mathbf{v}):f(\mathbf{k})\mapsto f(\mathbf{k}+k_i\tfrac{v_i}{c}\mathbf{e}_i)$, where $\mathbf{e}_i$ is the unit vector along the $i$th coordinate axis, $\mathbf{v}$ is the velocity vector of the scatterer, and $\mathbf{x}=(x_1,x_2,\dots)$ for any vector $\mathbf{x}$. In particular, we have $\hat{P}_{v}=\hat{S}_1(v\mathbf{e}_1)$. Now, let $\hat{\mathbf{L}}(\mathbf{v})$ be the $2n\times 2n$ matrix operator:
\begin{equation*}
\begin{bmatrix}
(1+\tfrac{v_1}{c})\hat{S}_1^{-1}(\mathbf{v}) & 0 & 0 & \cdots\\
0 & (1-\tfrac{v_1}{c})\hat{S}_1(\mathbf{v}) & 0 & \cdots\\
0 & 0 & (1+\tfrac{v_2}{c})\hat{S}_2^{-1}(\mathbf{v}) & \cdots\\
\vdots & \vdots & \vdots & \ddots
\end{bmatrix}\,.
\end{equation*}
Then, the transfer matrix for the scatterer moving with velocity $\mathbf{v}$ is given by
\begin{equation*}
\hat{\mathbf{L}}(\mathbf{-v})\,M\,\hat{\mathbf{L}}(\mathbf{v})\,,
\end{equation*}
where $M$ is the original transfer matrix for the scatterer, obtained in a manner such as that used to obtain~\eref{eq:M0}, for example. The ordering of the elements of $M$ is such that it acts on the vector $\big(A_1(\mathbf{k}),B_1(\mathbf{k}),A_2(\mathbf{k}),\dots\big)$:
\begin{equation*}
 \begin{pmatrix}
  C_1(\mathbf{k})\\
  D_1(\mathbf{k})\\
  C_2(\mathbf{k})\\
  \vdots
 \end{pmatrix}=\hat{\mathbf{L}}(\mathbf{-v})\,M\,\hat{\mathbf{L}}(\mathbf{v})\begin{pmatrix}
  A_1(\mathbf{k})\\
  B_1(\mathbf{k})\\
  A_2(\mathbf{k})\\
  \vdots
 \end{pmatrix}\,,
\end{equation*}
with $A_i(\mathbf{k})$ being the outgoing mode and $B_i(\mathbf{k})$ the incoming mode along the $i$th axis in the negative half-space (assuming that the scatterer is at the origin); and $C_i(\mathbf{k})$ the incoming mode and $D_i(\mathbf{k})$ the outgoing mode in the positive half-space.

\section{Quantum correlation function of the force operator\label{sec:Diffusion}}
In quantum theory, we need to replace the mode amplitudes $A(k)$ by operators $\hat A(k)$, and similarly for the $B$, $C$, and $D$ modes. 
The cross correlation of these operators is not trivial because of the boundary condition connecting the mode amplitudes $A(k)$, $B(k)$, $C(k)$ and $D(k)$. The input modes $\hat C(k)$ and $\hat B(k)$ can be considered independent, and the commutator is non-vanishing for the creation and annihilation operators of the same mode, e.g.,
\begin{align*}
\Big[ \hat B(k), {\hat B}^\dagger(k^\prime) \Big] & =  \left[ \hat C(k), {\hat C}^\dagger(k^\prime) \right] = \frac{\hbar \omega}{2 \epsilon_0 V} \delta_{k,k^\prime} \text{,}\\
\Big[ \hat B(k), {\hat C}^\dagger(k^\prime) \Big] & =0\,,
\end{align*}
assuming a discrete mode index of $k$, and a quantisation volume $V=S l$ with $S$ being the mode area and $l$ a fictitious total length of the space in one dimension. 

We consider only the $v=0$ case, since our expressions are accurate up to first order in $v/c$. In the quantum description, the linear relation for the output modes is
 \begin{align*} 
   A(k) = {}& \trans C(k) + \refl B(k) + \sqrt{\gamma} E \\
   D(k) = {}& \refl C(k) + \trans B(k) + \sqrt{\gamma} E\,,
\end{align*}
where the transmission $\trans=1/M_{22}=1/(1-i\zeta)$, and reflection $\refl=M_{12}/M_{22} = i\zeta/(1-i\zeta)$, as above. The fictitious amplitude $E$ represents the quantum noise fed into the system by the absorption. For $\gamma = 1-\big(|\refl|^2 + |\trans|^2\big)$, this noise ensures that the output modes obey the same commutation relations as the input ones, namely
\begin{align*}
\Big[ \hat A(k), {\hat A}^\dagger(k^\prime) \Big] &= \Big[ \hat D(k), {\hat D}^\dagger(k^\prime) \Big] = \frac{\hbar \omega}{2 \epsilon_0 V} \delta_{k,k^\prime}\,,\\
\Big[ \hat A(k), {\hat D}^\dagger(k^\prime) \Big] & =0\,.
\end{align*}
However, the linear dependence implies that commutators between input and output mode operators are
\begin{align*}
\Big[ \hat A(k), {\hat B}^\dagger(k^\prime) \Big] = \refl \Big[ \hat B(k), {\hat B}^\dagger(k^\prime) \Big] \,,\\
\Big[ \hat A(k), {\hat C}^\dagger(k^\prime) \Big] = \trans \Big[ \hat C(k), {\hat C}^\dagger(k^\prime) \Big] \,,
\end{align*}
and similar relations hold for the cross-commutators with $D(k)$.

The proper treatment of quantum fluctuations and the derivation of correlation functions require that the explicit time dependence be considered. Let us introduce the time-varying operators
\begin{equation*}
 \hat a(t) = \sum_k \hat A(k) e^{-i\omega t}\,,
\end{equation*}
and similarly for $ \hat b(t)$, $ \hat c(t)$ and $ \hat d(t)$. It follows that
\begin{equation*}
\Big[ \hat a(t), {\hat a}^\dagger (t^\prime) \Big] =  \frac{\hbar \omega}{2 \epsilon_0 V} \sum_k  e^{-i\omega (t-t^\prime)} \approx  \frac{\hbar \omega}{2 \epsilon_0 c S} \delta(t-t^\prime)\,.
\end{equation*}
Here we used that the non-excited, vacuum modes also contribute to force fluctuations. Therefore the Fourier-type summation extends to a broad frequency range and yields a $\delta(t-t^\prime)$ on the much slower timescale of interest. A similar commutation relation applies to the operators  $ \hat b(t)$, $ \hat c(t)$, and $ \hat d(t)$. The cross-commutators can be derived directly from those concerning the modes, e.g.,
\begin{equation*}
\Big[ \hat a(t), {\hat b}^\dagger (t^\prime) \Big] = \refl \frac{\hbar \omega}{2 \epsilon_0 c S}  \delta(t-t^\prime)\,.
\end{equation*}
The force operator is
\begin{equation}
\label{eq:ForceOp}
\hat \force = S\big[\hat \mst_{xx}({x\rightarrow 0^+})-\hat \mst_{xx}({x\rightarrow 0^-})\big]\,,
\end{equation}
as before, where
\begin{align*}
\hat \mst_{xx}(x\rightarrow 0) = \begin{cases}
-2\epsilon_0 \Big[\hat a^\dagger(t) \hat a(t) + \hat b^\dagger(t) \hat b(t)\Big]\vspace{0.5em}&x\rightarrow 0^-\\
-2\epsilon_0 \Big[\hat c^\dagger(t) \hat c(t) + \hat d^\dagger(t) \hat d(t)\Big]&x\rightarrow 0^+
\end{cases}
\end{align*}
is the quantized stress tensor. Assuming that the field is in a coherent state, in all normally ordered products, the mode amplitude operators can be replaced by the corresponding coherent state amplitudes, which are c-numbers: e.g.,\ $\hat A(k) \rightarrow A(k)$ and $\hat A^\dagger(k) \rightarrow A^\star(k)$. The force operator in \eref{eq:ForceOp} is normally ordered in this way; therefore coherent-state fields render, as a mean value of the quantum expressions, the force \eref{eq:MSTForce} derived from the classical theory based on the definition \eref{eq:Forcedef}. Non-trivial quantum effects arise from non-normally ordered products, such as the 4th-order product terms of the second order correlation function of the force \eref{eq:StartDiff}.  These terms can be evaluated straightforwardly by invoking the above-derived commutators to rearrange the product into normal order. As an example, consider
\begin{align*}
\big\langle \hat a^\dagger(t) \hat a(t) \hat a^\dagger(t^\prime) \hat a(t^\prime)\big\rangle =\ & \big\langle \hat a^\dagger(t) \hat a^\dagger(t^\prime) \hat a(t)  \hat a(t^\prime)\big\rangle \\ &-  \big\langle \hat a^\dagger(t)  \hat a(t^\prime)\big\rangle \frac{\hbar \omega}{2 \epsilon_0 c S} \delta(t-t^\prime)\,.
\end{align*}
For radiation fields in coherent state, the first term is canceled from the correlation function by the $\langle \hat a^\dagger(t)  \hat a(t^\prime)\rangle^2$ term. The coefficient of $\delta (t-t^\prime)$ in the second term is in normal order and can be replaced by c-numbers and then calculated identically as the force in \sref{sec:Force}, 
\begin{equation*}
 \label{eq:NormalOrderMean}
 \big\langle \hat a^\dagger(t)  \hat a(t)\big\rangle \approx \Big| \sum A (k) \Big|^2 = \frac{\hbar \omega}{2 \epsilon_0 c S} |A|^2\,,
\end{equation*}
in terms of the photo-current intensity $|A|^2$.

Assembling all similar contributions, originating from the non-vanishing commutators $[b,b^\dagger]$, $[c,c^\dagger]$, $[d,d^\dagger]$, $[a,b^\dagger]$, etc., one obtains~\eref{eq:DiffCoeff} presented in \sref{sec:MSTDiffusion}.

\section{Mirror cooling via the radiation pressure coupling Hamiltonian\label{sec:RadPressCool}}
We describe a generic opto-mechanical system composed of a single, damped-driven field mode coupled to the motion of a massive particle, whose Hamiltonian is given by~\cite{Courty2001,Vitali2003}
\begin{align*}
{\cal \hat{H}} =\ & \hbar \omega_c \hat{a}^\dagger \hat{a}  + i \hbar \eta (\hat{a}^\dagger e^{-i \omega t} - \hat{a} e^{i \omega t}) \\ & + \frac{\hat{p}^2}{2m} +V(\hat{x}) +\hbar G \hat{a}^\dagger \hat{a} \hat{x}\,.
\end{align*}
where $\hat{a}$ and $\hat{a}^\dagger$ are the annihilation and creation operators of the mode, $\hat{x}$ and $\hat{p}$ are the position and momentum operators associated with the motion and we drop the carets to signify expectation values. The mode is driven by a coherent field with an effective amplitude $\eta$ and frequency $\omega$.  This Hamiltonian describes, for example, the radiation pressure coupling of a  moving mirror to the field in a Fabry-Perot resonator. In this case the coupling constant is $G=\omega_c/L$, rendering the cavity mode frequency detuning $\omega_cx/L$ provided the mirror is shifted by an amount $x$. Since the cavity mode is lossy with a photon escape rate of $2 \kappa$, the total system is dissipative. Thereby, with a proper setting of the parameters, in particular the cavity detuning $\Delta_C=\omega-\omega_C$, the mirror motion can be cooled. We will determine the corresponding friction force linear in velocity.

In a frame rotating at frequency $\omega$, the Heisenberg equation of motion for the field mode amplitude reads  
\begin{equation*}
\dot{\hat{a}} = \left[ i (\Delta_C -G \hat{x}) - \kappa\right] \hat{a} + \eta\,.
\end{equation*}
where the noise term is omitted. We assume that the mirror moves along the trajectory $x(t) \approx x + v t$ with fixed velocity $v$ during the short time that is needed for the field mode to relax to its steady-state. The variation of $\hat{a}$ arises from the explicit time dependence and from the motion of the mirror.  A steady-state solution is sought in the form of $\hat{a} \approx \hat{a}^{(0)}(x) + v \hat{a}^{(1)}(x)$. On replacing this expansion into the above equation, and using the hydrodynamic derivative $\tfrac{d}{dt} \rightarrow \tfrac{\partial}{\partial t} + v \tfrac{\partial}{\partial x}$, one obtains a hierarchy of equations of different orders of the velocity $v$. To zeroth order the adiabatic field is obtained as
\begin{equation*}
a^{(0)} = \frac{\eta}{-i (\Delta_C-G x) + \kappa}\,.
\end{equation*}
The linear response of $a$ to the mirror motion is then
\begin{align*}
a^{(1)} &= \frac{1}{i (\Delta_C-G x) - \kappa} \frac{\partial}{\partial x} a^{(0)}\\
&=  \frac{i \eta G}{\big[-i ( \Delta_C-G x) + \kappa\big]^3}\,.
\end{align*}

The force acting on the mirror derives from the defining equation $\dot{\hat{p}}= \tfrac{i}{\hbar}[{\cal \hat{H}}, \hat{p}] =  -\hbar G \hat{a}^\dagger \hat{a}$. The force linear in velocity is
\begin{equation*}
\force_1 =  - 2v \hbar G \re{{a^{(0)\star}} a^{(1)}} =   4 v \frac{\hbar \eta^2 G^2 \kappa \Delta_C}{ \big[ \Delta_C^2 + \kappa^2\big]^3}\,,
\end{equation*}
where we used $x=0$ without loss of generality. It can be seen that mirror cooling requires that $\Delta_C<0$, \ie, the cavity resonance frequency is above the pump frequency. In this case, for efficient excitation of the field in the resonator, the frequency of the pump photons is up-shifted at the expense of the mirror's kinetic energy. This cooling force has been derived in \sref{sec:Mirrorcooling}, as a limiting case of the more general scattering theory. To check the perfect agreement between the two results,  the quantity corresponding to $\eta$ can be deduced from the total field energy in the resonator for an immobile mirror, which is $\hbar \omega_C {\hat{a}^{(0)\dagger}}\hat{a}^{(0)}$ here.

\end{document}